\documentclass{PoS}

\title{Contribution title}

\title{Update to the Bodek-Yang Unified Model for Electron- and Neutrino- Nucleon Scattering Cross Sections}

\ShortTitle{Update to the Bodek-Yang Unified Model for Electron- and Neutrino- Nucleon Scattering Cross Sections}

\author{\speaker{Arie Bodek}%
       University of Rochester\\
       E-mail: \email{bodek@pas.rochester.edu}}

\author{Un-Ki Yang\\
       University of Mancheter\\
             E-mail: \email{ukyang@cern.ch}}

\abstract{We construct a  model for inelastic neutrino- and electron-nucleon 
scattering cross sections using effective leading order parton distribution
functions with a new scaling variable $\xi_w$. 
Non-perturbative effects 
are  well described using the  $\xi_w$ scaling variable, in combination
with multiplicative $K$ factors at low $Q^2$.
Our model describes all inelastic charged lepton-nucleon scattering
(including resonance) data (HERA/NMC/BCDMS/SLAC/JLab) ranging from very high $Q^2$
 to very low $Q^2$ and down to the  photo-production region. The model describes existing
  inelastic  neutrino-nucleon scattering measurements, and
is currently used in analyses of  neutrino oscillation experiments
 in the few GeV region.}

\FullConference{35th International Conference of High Energy Physics - ICHEP2010,\\
		July 22-28, 2010\\
		Paris France}

\begin{document}


Standard  PDFs are extracted  from global fits to various sets of
deep inelastic (DIS) scattering data 
at  high energies and high $Q^2$, where non-perturbative QCD effects are negligible.
PDF fits are performed within the framework of QCD in either  LO, NLO or  NNLO.

In order to use the PDFs at low $Q^2$ we use a  new scaling variable ($\xi_w$)  to  construct  effective LO PDFs 
that  account for the contributions from  target mass corrections,
 non-perturbative QCD effects,  and higher order QCD terms.


Our proposed scaling variable, $\xi_w$ is derived as follows.
Using energy momentum conservation, the fractional momentum, $\xi$ 
carried by a quark in a proton target of mass $M$ 
is 
\begin{eqnarray}
 \xi &=& \frac{2xQ^{'2}}{Q^{2}(1+\sqrt{1+(2Mx)^{2}/Q^2})} \nonumber \\
2Q^{'2} & =& [Q^2+M_f{^2}-M_i{^2}] + \sqrt{(Q^2 + M_f{^2}-M_i{^2})^2+4Q^2(M_i{^2}+P_{T}^{2})}\nonumber 
\end{eqnarray}

Here $M_i$ is the initial quark mass with average initial
transverse momentum $P_T$,  and $M_f$ is the mass of the final state 
quark.  
Assuming $M_i=0, P_T=0$ we construct  following scaling variable
\begin{eqnarray}
\label{eq:xi}
\xi_w &=& \frac{2x(Q^2+M_f{^2}+B)}
        {Q^{2} [1+\sqrt{1+(2Mx)^2/Q^2}]+2Ax},\nonumber
\end{eqnarray}
where in general $M_f =0$ (except  for the
case of charm-production in neutrino scattering  for which $M_f$=1.32 GeV).
The parameter $A$ is used to
account (on average)  for the higher order QCD terms and dynamic higher twist  
in the form of an enhanced target mass term (the effects of the proton target 
mass is already taken into account  in the denominator of $\xi_w$).
 The parameter
$B$ is used to account (on average) for the initial state quark transverse
momentum,  and also for the effective  mass of the final state quark 
originating from multi-gluon emission. 
A  non-zero $B$ also  allows us to describe data in
the photoproduction limit (all the way down to $Q^{2}$=0).

In leading order QCD (e.g.
GRV98 PDFs), ${\cal F}_{2,LO}$ for the scattering
of electrons and muons on proton (or neutron) targets is
given by the sum of quark
and anti-quark distributions (each weighted the
square of the quark charges):
\begin{eqnarray}
{\cal F}_{2, LO}^{e/\mu}(x,Q^{2}) = \Sigma_i e_i^2 \left [xq_i(x,Q^{2})+x\overline{q}_i(x,Q^{2}) \right].\nonumber
\end{eqnarray}
Our proposed effective LO PDFs model includes the following: 
\begin{enumerate}
 \item The GRV98 LO Parton Distribution Functions (PDFs) 
   are used to describe  ${\cal F}_{2, LO}^{e/\mu}(x,Q^{2})$.
	 The minimum $Q^2$ value for these PDFs is 0.8 GeV$^2$.
 \item  The scaling variable $x$ is replaced with the 
        scaling variable $\xi_w$.  
        \begin{eqnarray}
      {\cal F}_{2, LO}^{e/\mu}(x,Q^{2})  =  \Sigma_i e_i^2 
       \left [\xi_wq_i(\xi_w,Q^{2})+\xi_w\overline{q}_i(\xi_w,Q^{2}) \right].\nonumber
        \end{eqnarray}
\item  
	We multiply all PDFs by vector $K$ factors such that they have the
	correct form in the low $Q^2$ photo-production limit. We 
	use different forms for the sea and valence quarks.  
      $$ K_{sea}^{vector}(Q^2) = \frac{Q^2}{Q^2 +C_s} ,~~~ 
	 K_{valence}^{vector}(Q^2) =[1-G_D^2(Q^2)]
	      	\frac{Q^2+C_{v2}} 
		      {Q^{2} +C_{v1}}$$ 
	 where $G_D$ = $1/(1+Q^2/0.71)^2$ is the  proton elastic form factor.
	 This form for the $K$ factor for valence quarks is motivated
	  by closure arguments
	  and the Adler 
	  sum rule. 
	At low $Q^2$, $[1-G_D^2(Q^{2})]$
	is approximately $Q^2/(Q^2 +0.178)$.	
        These modifications are included  in order to describe low $Q^2$
       data in the photoproduction limit ($Q^2$=0), where	
       ${\cal F}_{2}^{e/\mu}(x,Q^2)$ is related to the photoproduction cross section 
	according to
	\begin{eqnarray}
	     \sigma(\gamma p) = {4\pi^{2}\alpha_{\rm EM}\over {Q^{2}}}
	          {\cal F}_{2}^{e/\mu}(x,Q^2)
	           = \frac{0.112 mb}{Q^2} {\cal F}_{2}^{e/\mu}(x,Q^2)\nonumber
	\end{eqnarray}
 \item We freeze the evolution of the GRV98 PDFs at a
	value of $Q^2=0.80$ GeV$^2$. Below this $Q^2$, ${\cal F}_2$ is given by;
	\begin{eqnarray}
	     {\cal F}_2^{e/\mu}(x,Q^2<0.8) =K^{vector}(Q^2)
		 {\cal F}_{2,LO}^{e/\mu}(\xi_{w},Q^2=0.8) \nonumber
	\end{eqnarray}

 \item Finally, we fit for  the parameters  of the modified
   effective GRV98 LO PDFs (e.g. $\xi_w$) 
       to  inelastic   charged lepton scattering
        data  on hydrogen and deuterium targets 
         (SLAC, BCDMS, NMC, H1
         ).
        We obtain an excellent fit with the following initial parameters: 
        $A$=0.419, $B$=0.223, and  $C_{v1}$=0.544, $C_{v2}$=0.431,
        and $C_{sea}$=0.380, with  $\chi^{2}/DOF=$ 1235/1200.
	Because of these additional  $K$ factors, we find that
        the GRV98 PDFs need to be scaled up by a normalization
        factor  $N$=1.011.  
   \end{enumerate}

We now describe the second iteration of the fit. 
Theoretically, the $K_{i}$ factors  
are not required
to be the same for the $u$ and $d$  valence quarks or the $u$,  $d$  and $s$
 sea quarks and antiquarks.
In order to allow flexibility in  our effective LO model, we treat the  $K_{i}$ factors 
for $u$ and $d$ valence  and sea quarks separately.
As the predictions of our  model are in good agreement
with photoproduction data, and 
for much of the resonance region,
 we now  proceed to include photo-production data  in the fit.  In order
to get  additional constraints on the different  $K$ factors for up
and down quarks separately, we  use both hydrogen and deuterium data.

 The second iteration   
includes  the additional photo-production and resonance
data in the fit,.

$$ K^{LW}  =\frac{\nu^2+ C^{L-Ehad}  } {\nu^2}, ~~
		 K_{sea-strange}^{vector}(Q^2) = \frac{Q^2}{Q^2 +C_{sea-strange}^{vector}}, ~
	 	 K_{sea-up}^{vector}(Q^2) = \frac{Q^2}{Q^2 +C_{sea-up}^{vector}},$$
	 $$	 K_{sea-down}^{vector}(Q^2) = \frac{Q^2}{Q^2 +C_{sea-down}^{vector}}, ~~~
	 K_{valence-up}^{vector}(Q^2)=K^{LW}[1-G_D^2(Q^2)]
	     \frac{Q^2+C_{v2u}^{vector}} {Q^{2} +C_{v1u}^{vector}}, $$
		   $$   K_{valence-down}^{vector}(Q^2) =K^{LW}[1-G_D^2(Q^2)]
	   \frac{Q^2+C_{v2d}^{vector}} 
		      {Q^{2} +C_{v1d}^{vector}} $$	
	The best fit is given by 
  $A=0.621 \pm 0.009$, $B=0.380 \pm 0.004$, $C_{v1d}^{vector}=0.341 \pm 0.007$,
$C_{v1u}^{vector}=0.417  \pm 0.024$, $C_{v2d}^{vector}=0.323 \pm 0.051$,  $C_{v2u}^{vector}=0.264 \pm 0.015$, and 
$C^{L-Ehad}=0.217 \pm 0.015$. The sea
factors are  $C_{sea-down}^{vector}$=0.621, $C_{sea-up}^{vector}$=0.363,  and $C_{sea-strange}^{vector}$ was set to be the same as $C_{sea-down}^{vector}$ . Here,  the parameters are in units of $(GeV/c)^2$.   
The fit  $\chi^{2}/DOF=$2357/1717, and   $N=1.026 \pm 0.003$. 
The resonance data add  to the $\chi^2/ndf$  because the fit only
provides a smooth average over the resonances. No neutrino data
are included in the fit.

For high energy neutrino scattering on quarks
and antiquarks,  the vector and axial
contributions are the same.  
At  very high $Q^2$,  where the quark
parton model is valid, both  the vector and axial  $K$ factors 
expected to be 1.0. Therefore 
 neutrinos and antineutrino structure
 functions for
 an iso-scalar target are given by :
\begin{eqnarray}
{\cal F}_{2}^\nu(x,Q^{2}) &=& \Sigma_i 2\left [\xi_w q_i(\xi_w,Q^2) +\xi_w \overline{q}_i(\xi_w,Q^2)  \right].\nonumber \\
x{\cal F}_{3}^\nu(x,Q^{2}) &=&  \Sigma_i 2\left [\xi_w q_i(\xi_w,Q^2) -\xi_w \overline{q}_i(\xi_w,Q^2) \right]. \nonumber  
\end{eqnarray}

There are two major differences between the case of  electron/muon
 inelastic scattering and the case of
 neutrino and antineutrino scattering.
 In the neutrino case we have one additional
 structure function  ${\cal F}_{3}^\nu(x,Q^{2})$. 
 In addition, at  low $Q^{2} $ 
there should be a difference between
the vector and axial $K_{i}$ factors due a difference in the 
non-perturbative  axial vector contributions.
Unlike the vector ${\cal F}_2$ which must go to zero
in  the $Q^2=0$ limit, the axial part of ${\cal F}_2$ is 
non-zero in  the $Q^2=0$ limit.

We account for kinematic and dynamic  higher twist and higher order 
QCD effects in ${\cal F}_{2}$  by fitting  
 the  parameters of the scaling variable $\xi_w$ 
 and the  $K$ factors 
 to low $Q^2$ data for  ${\cal F}_{2}^{e/\mu}(x,Q^{2})$.  These
 should also be valid the vector part of ${\cal F}_2$ 
 in neutrino scattering. 
 However,
the higher order QCD effects in  ${\cal F}_{2}$ and $x{\cal F}_{3}$
are different.  We account for the different scaling violations
in  ${\cal F}_{2}$ and $x{\cal F}_{3}$
by adding another correction $H(x,Q^{2})$ to the
expression for $x{\cal F}_{3}$

 The differences between neutrinos and
 charged lepton scattering 
are accounted for in the following expressions:

$${\cal F}_{2}^{\nu vector}(x,Q^{2}) =
  \Sigma_i K_i^{vector}(Q^2) \xi_w q_i(\xi_w,Q^2)
 +  \Sigma_j K_j^{vector}(Q^2) \xi_w \overline{q}_j(\xi_w,Q^2)$$
$$
{\cal F}_{2}^{\nu axial}(x,Q^{2})=
 \Sigma_iK_i^{axial}(Q^2) \xi_w q_i(\xi_w,Q^2)
  +      \Sigma_j K_j^{axial}(Q^2)   \xi_w \overline{q}_j(\xi_w,Q^2)$$  
 $$
x{\cal F}_{3}^\nu(x,Q^{2} ) =  2  H(x,Q^{2}) 
\left [ \Sigma_i K^{xF3}_i   \xi_w q_i(\xi_w,Q^2)  
 -  \Sigma_j K^{xF3}_j   \xi_w \overline{ q}_j(\xi_w,Q^2)
 \right]$$
Where $i$ denotes $(valence-up)$, $(valence-down)$, $(sea-up)$,
$(sea-down)$,  and $(sea-strange)$.
Detailed expressions are given in reference\cite{bypaper}.

With
 the above assumptions we
 calculate the differential cross sections for
 neutrinos and antineutrino scattering. 
We also  correct for nuclear effects in iron
using the ratio of iron to deuterium structure functions
as measured in muon and electron scattering experiments.

 Our predictions are in  good agreement with the 
CCFR 
CDHSW 
neutrino and antineutrino differential cross sections.


\begin{thebibliography}{99}
\bibitem{bypaper} A. Bodek and U.K. Yang,  hep-ph/1011.6592.


\end{thebibliography}
\end{document}